\input harvmac
\input psfig
\newcount\figno
\figno=0
\def\fig#1#2#3{
\par\begingroup\parindent=0pt\leftskip=1cm\rightskip=1cm\parindent=0pt
\global\advance\figno by 1
\midinsert
\epsfxsize=#3
\centerline{\epsfbox{#2}}
\vskip 12pt
{\bf Fig. \the\figno:} #1\par
\endinsert\endgroup\par
}
\def\figlabel#1{\xdef#1{\the\figno}}
\def\encadremath#1{\vbox{\hrule\hbox{\vrule\kern8pt\vbox{\kern8pt
\hbox{$\displaystyle #1$}\kern8pt}
\kern8pt\vrule}\hrule}}
\def\underarrow#1{\vbox{\ialign{##\crcr$\hfil\displaystyle
 {#1}\hfil$\crcr\noalign{\kern1pt\nointerlineskip}$\longrightarrow$\crcr}}}
% use of underarrow
%A~~~\underarrow{a}~~~B
%
\overfullrule=0pt

%macros
%

%

\font\zfont = cmss10 %scaled \magstep1

\def\bigone{\hbox{1\kern -.23em {\rm l}}}
\def\ZZ{\hbox{\zfont Z\kern-.4emZ}}

\Title{hep-ph/0002297}
{\vbox{\centerline{The Cosmological Constant}
\bigskip
\centerline{From The Viewpoint Of String Theory}}}
\smallskip
\centerline{Edward Witten$^*$}
\smallskip
\centerline{\it Dept. of Physics, Cal Tech, Pasadena, CA 91125}
\smallskip\centerline{and}
\smallskip
\centerline{\it CIT-USC Center For
Theoretical Physics, USC, Los Angeles CA}

\medskip

\noindent
The mystery of the cosmological constant is probably the most pressing
obstacle to significantly improving the models of elementary particle
physics derived from string theory.  The problem arises because
in the standard framework of low energy physics, there appears to be
no natural explanation for vanishing or extreme smallness of the vacuum
energy, while on the other hand it is very difficult to modify this
framework in a sensible way.  In seeking to resolve this problem,
one naturally wonders if the real world can somehow be interpreted
in terms of a vacuum state with unbroken supersymmetry.
\vskip 3cm
\noindent 
Lecture at DM2000, Marina del Rey, February 23, 2000.
\vskip .5cm
\noindent
$^*$ On leave from Institute for Advanced Study, Princeton NJ 08540.
\vskip 1cm
\Date{March, 2000}
%text of paper

The problem of the vacuum energy density or
cosmological constant -- why it is zero or extremely
small by particle physics standards -- really only arises in the presence
of gravity, since without gravity, we don't care about the energy of the
vacuum.  Moreover, it is mainly a question about quantum gravity, since
classically it would be more or less natural to just decide -- as Einstein
did -- that we do not like the cosmological constant, and set it to zero.

Classically, it may involve some fine-tuning to set the cosmological
constant to zero, but once this is done, that is the end of the story.
Quantum mechanically, it will not help to just set the cosmological
constant to zero in a microscopic Lagrangian.  The observed value of the
vacuum energy density potentially includes contributions from things
such as loops of soft photons.  The puzzle is why the vacuum energy
is so small after including all of these contributions.

As the problem really involves quantum gravity, 
string theory is the only framework for addressing it, at least
with our present state of knowledge.
Moreover, in string theory, the question is very sharply posed, as there 
is no dimensionless parameter.  Assuming that the dynamics gives a unique
answer for the vacuum, there will be a unique prediction for the cosmological
constant.

But that is, at best, a futuristic way of putting things.  We are not
anywhere near, in practice, to understanding how there would be a 
unique solution for the dynamics.  In fact, with what we presently
know, it seems almost impossible for this to be true, rather as a few
years ago it seemed almost impossible that the different string theories
would turn out -- as they have -- to be limiting cases of one more
unified theory.

Not understanding why the cosmological constant is zero, or extremely
small, is in my opinion the key obstacle to making the models of
particle physics that can be derived from string theory more realistic.
In fact, as I will explain, that has been true since 1985.

Any time that one does not understand something, one can point to details
that do not work.   It is always important
to identity what is wrong qualitatively and gives the best clue to possible
future progress.  

Let me give an analogy.  In the early 1980's, it ws fairly clear that string
theory gave consistent models of quantum gravity and that in that 
framework, one was forced to unify gravity with  other forces.
Many details were not right, but the most striking qualitative problem
was that in the models of particle physics that could be derived from
string theory at that time {\it the weak interactions would have to conserve
parity}.  Parity violation in weak interactions is deeply embedded in the
standard model, and the inability to reproduce it is a much more basic
problem than an inability to compute the correct mass of the muon,
for example.  

When this problem was cleared up, by the Green-Schwarz generalized
anomaly cancellation mechanism in 1984, many other things fell into
place -- including the discovery of the heterotic string shortly
afterwards -- and the particle physics models became much more realistic.

Since that time, the outstanding qualitative problem has been the cosmological
constant, or more exactly, why it is zero or extremely small after 
supersymmetry breaking.  Broadly speaking, models of particle physics
derived from string theory work fine in the absence of supersymmetry breaking;
but we do not have a convincing method of supersymmetry breaking.
Hence, we can give a reasonably elegant (though not rigorous)
explanation of quark and lepton gauge interactions and family
structure.  But we cannot say very much about masses or mixing angles.

The basic diagnostic test for knowing that a wrong mechanism of
supersymmetry breaking is wrong is that it generates a cosmological
constant.  On this basis, all the known approaches to supersymmetry breaking
are wrong.  Thus, the cosmological constant is the main clue to improving
the particle physics models.

Without supersymmetry breaking, there is no problem in getting a stable 
vacuum with zero cosmological constant.  If $\phi$ is a generic
scalar field and $V(\phi)$ is the effective potential, then with
unbroken supersymmetry one can naturally have a stable minimum of $V(\phi)$
at which $V=0$.  (In the abstract, unbroken supersymmetry allows
negative $V$ as well, but when one actually tries to construct string
theory models that could reproduce the supersymmetric standard model 
-- for example via Calabi-Yau compactification -- one naturally
finds $V=0$.)

Known mechanisms of supersymmetry breaking typically give
an unstable runaway, with $V$ positive but vanishing as $\phi$ goes
to infinity, or else they give
 a $V$ that is not positive definite.  (The runaway
involves one or more scalars, depending on the model considered.
I will use ``$\phi$'' as a generic label for such scalars.)
This is linked to the fact that  the mechanisms of supersymmetry
breaking
that we understand generally turn off when the string coupling constant
goes to zero or a compactification radius goes to infinity.
And there generally is a scalar field $\phi$ that controls
the string coupling constant or compactification radius.
So supersymmetry breaking turns off as $\phi$ goes to infinity 
\ref\ds{M. Dine and N. Seiberg,
``Is The Superstring Semiclassical?'' in {\it Unified String Theories},
ed. M. B. Green and D. J. Gross (World Scientific, 1986),
``Is The Superstring Weakly Coupled?'' Phys. Lett. {\bf B156}
(1985) 55.}.

Scenarios with an unstable runaway have been considered by
many physicists over the years.  As far as I know, such scenarios
were first considered by Dirac in his approach to the large numbers
problem.  In recent years, experimental   evidence for a nonvanishing cosmic
energy density in today's universe has suggested that (as an
alternative to a true cosmological ``constant'') the actual
universe may be undergoing such a runaway.  Such scenarios
have been much discussed (along with other possible cosmic components
of negative pressure) under the name ``quintessence.''   A discussion
of the observational situation can be found in \ref\wcos{L. Wang,
R. R. Caldwell, J. P. Ostriker, and P. J. Steinhardt, ``Cosmic Concordance
and Quintessence,'' astro-ph/9901388.}.  These issues will be further
addressed by other speakers at the present meeting.

Going back to string theory, in view of the remarks in the last
paragraph, 
supersymmetry breaking mechanisms that lead to an unstable runaway
-- $V(\phi)$ has its minimum at $\phi=\infty$, where it vanishes --
should be considered seriously.  But there are severe difficulties
with them.  First of all, as $\phi\to\infty$, we need the bose-fermi
mass splittings to vanish fast enough relative to the vanishing of
$V(\phi)$.  This is
simply the cosmological constant problem restated for the case of 
a runaway solution rather than a stable vacuum.  Known supersymmetry
breaking mechanisms fail this test.  

Moreover, even if we can arrange for 
$V(\phi)$ to vanish fast enough for large
$\phi$, compared to the mass splittings, this kind of scenario
has to face acute obstacles.  Unless we restrict the couplings
of $\phi$ in a way that has no apparent rationale in mechanisms of
supersymmetry breaking that I can think of, we will find that as 
$\phi$ varies, the natural ``constants'' will change in time.
We are at risk of finding  $\dot G/G$ and 
$\dot\alpha/\alpha$ close to $\dot \phi/\phi$
(here $G$ is Newton's constant, measured in units in which $\hbar,$
$c$, and the proton mass are fixed; $\alpha$ is the fine structure
constant).  By making $V$ sufficiently flat so that
 $\dot\phi/\phi$ is sufficiently small, and making the couplings
of $\phi$ sufficiently weak, we can perhaps bring $\dot G/G$ and 
$\dot \alpha/\alpha$ within acceptable limits.  

Even so, we will still, in general, be in trouble because
if $V$ is so flat and $\dot \phi$ so small as such a scenario
will require, then $\phi$ will behave in laboratory and solar system
measurements as a massless field.
 We will generically expect that coherent and perhaps even
spin-dependent forces mediated by $\phi$ will show up in
tests of the   equivalence principle, solar system tests of General
Relativity, and perhaps in ``fifth force'' measurements.  
I actually personally find it very surprising, if nature is exhibiting
a runaway with cosmic evolution of a scalar field $\phi$, that
the effects of the $\phi$ field have not been seen directly already
in one or more of the experiments that I have mentioned.

Especially if experimental indications of a cosmological
``constant'' hold up,  I think that experiments that will improve the
bounds on $\phi$ couplings are important and promising.  One
proposal, for example, is a satellite measurement (STEP) 
that could possibly
improve the tests of the equivalence principle by a factor of $10^6$.
This would improve the bounds on the couplings of $\phi$ by a factor
of $10^3$. The bound at present is roughly speaking that the couplings of
$\phi$ are a couple orders of magnitude smaller than gravitational
coupling.  In string theory, a $\phi$ field typically has a coupling
not much weaker than gravity, and suppression of the $\phi$ couplings
relative to gravity puts a severe  restriction on the model.  From
this perspective,  in runaway scenarios
 it would be surprising if $\phi$ exists and would not be 
detected  in an experiment that would improve the test of the equivalence
principle by a factor of $10^6$.

I have emphasized scenarios with a runaway because I think that
they are suggested by experimental findings of possible
cosmic acceleration.   One advantage
of a runaway rather than a true cosmological ``constant'' is that,
by analogy with a zero cosmological constant
scenario first outlined by Dyson many years ago
\nref\dyson{F. W. Dyson, Rev. Mod. Phys. {\bf 51} (1979) 447.}%
\nref\frautschi{S. Frautschi, ``Entropy In An Expanding Universe,''
Science {\bf 217} (1982) 593.}%
\refs{\dyson,\frautschi}, in a runaway scenario, life can possibly
adapt and survive and develop forever
 by working at lower and lower temperatures and
with longer and longer time and length scales.  A strict cosmological constant
would bring all this to a grim end by introducing effective length
and time cutoffs.  On the other hand, as I  have also
tried to explain, the difficulties
with runaway scenarios are so severe that one would probably
have to find something
really interesting and new to get a scenario that works.

I should also mention an alternative (which still fits within
the general rubric of ``quintessence'') which is less ambitious and
dramatic than the runaway but faces much less severe experimental
difficulties.  We could assume that the potential $V(\phi)$ has a stable
minimum at which it vanishes, but that the actual universe has not
yet reached the minimum because the potential is very flat.  For
example, $\phi$ might be an axion-like field, which in particular
is angle-valued.  The potential, coming from instantons, might
be something like $V(\phi)=V_0(1-\cos\phi)$, with $V_0$ a constant.  I have
adjusted an additive constant in $V$ so that the true minimum of the
potential (at $\phi=0$) has $V=0$ and zero cosmological constant.
Either because $V_0$ is very small or the universe started
near the top of the potential bump at $\phi=\pi$
(or both), $\phi$ might still in today's universe be away
from the minimum of the potential and very slowly changing so that
it imitates a cosmological constant.
Axion-like fields that get exponentially
small potentials of roughly this kind do occur in many string models.
If the universe is near the maximum of the potential today, or at least
not too close to the minimum, then $V_0$ should be of order the 
observed vacuum energy density.  {\it A priori}, this involves
some fine-tuning, as does any description of a tiny but
nonzero vacuum energy density.  Such a scenario is not as challenging
conceptually as a runaway.  However, it has the practical advantage
that there are mechanisms for suppressing the axion couplings to ordinary
matter that would not apply to moduli that might lead to a runaway.
Hence, in this kind of scenario, the experimental limits on light
scalars are potentially much less problematic.  In such a scenario,
theorists would need to focus on why the true cosmological constant
is exactly zero and why there is an axion with the right value for $V_0$. 

What about other scenarios?  Interestingly, I cannot think of
any known assumption or approximation in string theory that leads to 
the most obvious interpretation of recent experimental findings:
a stable vacuum with a positive cosmological constant.  
There certainly is not an attractive known way to do this,
and  we certainly don't have the techniques at the moment
to find such a vacuum in which, in addition, the cosmological constant
is extremely small and the particle physics is attractive.

We do have stable vacua with negative cosmological constant
(and sometimes unbroken supersymmetry).  Such vacua presumably
do not describe the real world, but exploring them has been
in the last few years, beginning with Maldacena's proposed duality
\ref\malda{J. Maldacena, ``The Large $N$ Limit Of Superconformal Field
Theories And Supergravity,'' Adv. Theor. Math. Phys. {\bf 2} (1998) 231.},
an arena for amazing theoretical advances.  From this
work has come new approaches to nonperturbative
description of quantum gravity, understanding of quark confinement
in terms of properties of black holes, and more.

What I have said so far has been very general.
In the concluding portion of this talk, I will make a few remarks
on some specific approaches to the cosmological constant problem.
First of all, the problem is hard because of the following set of
facts:

(1) Solving the problem seems to require a low energy mechanism -- to
cancel contributions of loops of soft photons, for example.

(2) But low energy physics in the standard framework of four-dimensional
effective
field theory does not seem to offer a solution to the problem.
For a review of these issues, see \ref\weinberg{S. Weinberg, ``The
Cosmological Constant Problem,'' Rev. Mod. Phys. {\bf 61} (1989) 1.}.

(3) It is very hard to change the low energy framework in a sensible
way, given all of the familiar successes.

Faced with this conundrum, one way out (eloquently described by 
Weinberg in the lecture before mine) is an anthropic principle.
This entails abandoning the quest for a conventional scientific 
explanation and interpreting the smallness of the cosmological constant
as a necessary feature of our local environment, without which
our human species could not have developed to ponder the question.
I very much hope that things will not go that way, for several reasons.
First, I think we need the cosmological constant as a clue to understand
particle physics better; if the cosmological constant must be treated
anthropically -- and the approximate vacuum we live in is
drawn fortuitously from an ensemble -- I am not optimistic about how
well we will be able to understand the aspects of particle physics
that are not explained by the standard model.  Second, I think we all
prefer to see a conventional scientific explanation
because it gives more understanding.  Time and again the quest for
a conventional scientific explanation has triumphed over seemingly
impossible obstacles.  And finally, I want to ultimately
understand that, with all the particle physics one day worked out,
life is possible in the universe because $\pi$ is between 3.14159
and 3.1416.  To me, understanding this would be the real anthropic
principle.  I don't want to lose it.

On the other hand, it is not clear to me which alternative
approaches to the
cosmological constant are actually worth mentioning in the ten minutes
remaining.  I will make a slightly eccentric choice, guided by the
following considerations. 

Supersymmetry, if unbroken, would give a natural explanation of
the vanishing of the cosmological constant.\foot{This is not automatic,
since unbroken supersymmetry can coexist with negative cosmological
constant, as I have mentioned earlier.  But in many situations,
with higher dimensions, chiral symmetries, natural microscopic constructions
via Calabi-Yau manifolds, etc.,  unbroken supersymmetry does make
vanishing of the cosmological constant natural.}
It seems a pity to waste it.

\nref\wit{E. Witten, ``Instability Of The Kaluza-Klein Vacuum,'' Nuc. Phys.
{\bf B195} (1982) 481.}%
\nref\hor{M. Fabinger and P. Horava, ``Casimir Effect Between World
Branes In Heterotic $M$ Theory,'' hep-th/0002073.}
Moreover, unbroken supersymmetry, if valid, give a natural explanation
of the stability of spacetime.  This last point perhaps
needs some explanation.  The Einstein action  $\int d^4x\sqrt g R$ 
has no obvious positivity or stability property.  Positive energy
of classical General Relativity can nonetheless be proved -- for
instance, using the possibility of extending the classical theory
to include fermions in a supersymmetric way.  But positive energy
and stability of the vacuum fail in many plausible extension of General
Relativity such as nonsupersymmetric Kaluza-Klein theory
\refs{\wit,\hor}.  I think it is a mystery why any nonsupersymmetric
string vacuum would be stable.  Hence, I wonder if we can somehow
make use of the vacua  with unbroken supersymmetry to describe nature.

Moreover, a decade ago, one might have hoped that the supersymmetric
string vacua were nonperturbatively inconsistent.  It has become
pretty clear with all the results of the 1990's on nonperturbative
string dualities that this is not so.  If they are consistent, it
is a shame to waste them.

Can we somehow reinterpret the real world in terms of {\it unbroken}
supersymmetry, suitably construed, even though the
boson and fermion masses are different?

I know of two tries in this direction.  Both involve relating the
observed $D=4$ universe to a world of a different dimension.
One scenario starts with $D<4$, and the other starts with $D>4$.

For the first scenario \ref\witten{E. Witten, ``Is Supersymmetry
Really Broken?'' Int. J. Mod. Phys. {\bf A10} (1995) 1247, hep-th/9409111,
``Some Comments On String Dynamics,'' in {\it Strings '95: Future
Perspectives In String Theory}, ed. I. Bars et. al., hep-th/9507121.},
we consider a supersymmetric string vacuum in $D=3$ (or similarly
in $D<3$) with the dilaton or string coupling constant $g$ as the only
modulus.  Despite the supersymmetry of the vacuum, because of a peculiar three-dimensional infrared 
divergence, the boson and fermion masses are not equal -- they are equal
at tree level, but are split by gravitational corrections.

We assume that, for example because of a discrete $R$-symmetry, 
the cosmological constant vanishes.  Now we take $g\to \infty$.
We have learned in the 1990's that sometimes when a coupling constant
goes to infinity, a new dimension of spacetime opens up.  Let us assume
this occurs in the present case.  The cosmological constant will remain
zero because it is zero for all $g$.  Conceivably, as $g\to \infty$,
the boson and fermion masses (which were unequal for all finite nonzero $g$)
remain unequal. To know, we would need some new insight about
dynamics.  If so, the $g\to\infty$ limit would be a four-dimensional
world with vanishing cosmological constant and unequal bose-fermi masses.
It is not obvious that it could be interpreted, in a dual description,
as a four-dimensional world with conventional spontaneously broken
supersymmetry, but perhaps this is also possible.

Are there any models that actually exhibit the rather optimistic
dynamics that I have just suggested?  One reason that it is hard
to be find out is that models whose nonperturbative dynamics we know
something about usually have many moduli and can undergo partial
or complete decompactification even for weak coupling.  Because of the
way the scenario depends on an infrared divergence that is limited
to $D\leq 3$, I think that it is unlikely that the desired dynamics
will occur in any model that can undergo 
partial decompactification to $D\geq 4$ at fixed coupling.  
This will make it hard to learn in the near future
with known methods if the  scenario does work as hoped.

Now, I will discuss an alternative idea, based on a suggestion
by Gregory, Rubakov, and Sibiryakov
\ref\grs{R. Gregory, V. A. Rubakov, and S. M. Sibiryakov,
``Opening Up Extra Dimensions At Ultra Large Scales,'' hep-th/0002072.}.
They proposed a scenario in which the world looks four-dimensional
up to a very large (presumably cosmological) length scale $R$, but
 above that length scale is $D$-dimensional with $D\geq 5$.  
A discussion of the model aiming to verify that it has the
claimed quasi four-dimensional behavior can be
found in \ref\csaki{C. Csaki, J. Erlich, and T. J. Hollowood,
``Quasi-Localization Of Gravity By Resonant Modes,'' hep-th/0002161.}.

Since the GRS scenario does not obey the usual assumptions of
four-dimensional low energy effective field theory, which lead to the
cosmological constant problem, we should naturally reexamine the problem
in this scenario.  (In fact, this was done independently in a paper
\ref\gaba{G. Dvali, G. Gababadze, and M. Porrati,
``Metastable Gravitons And Infinite Volume Extra Dimensions,''
hep-th/0002190.} 
that appeared on the hep-th bulletin board the same day this
talk was given.  In this paper, it is also claimed that the GRS scenario
leads to scalar-tensor gravity rather than pure tensor and so is
excluded experimentally.)

For our purposes, we will start with a world of $D\geq 5$ with
unbroken supersymmetry.  For example, to be definite, we will take
$D=5$.  We assume that the five-dimensional cosmological constant
vanishes.  (This is natural in some supergravity models in five dimensions,
and would, according to Nahm's theorem, automatically occur in all
supersymmetric models in $D\geq 8$.)  We assume that we live on
a four-dimensional ``brane'' which is nonsupersymmetric (that is,
non-BPS).  This assumption makes the observed macroscopic four-dimensional
physics nonsupersymmetric \ref\hlp{J. Hughes, J. Liu, and J. Polchinski,
``Supermembranes,'' Phys. Lett. {\bf B180} (1986) 370}.  
It is natural for the brane to be flat because this gives a minimal
volume hypersurface in five-dimensional Minkowski space.

So we naturally get a flat four-dimensional world with unequal
masses for bosons and fermions.  But do we observe four-dimensional
gravity?  The idea of GRS was to get approximate four-dimensional
gravity by ``localizing'' a graviton on the brane along the lines
of Randall and Sundrum \ref\rs{L. Randall and  Sundrum,
``An Alternative To Compactification,'' Phys. Rev. Lett. {\bf 83} (1999) 4690,
hep-th/9906064.}.
The GRS idea can be illustrated with a metric of the general form
\eqn\ulmigu{ds^2=dr^2+\left(e^{-2r/r_0}+\epsilon\right)\sum_{i=1}^4 dx_idx^i.}
Here $r\geq 0$; there is some sort of ``brane'' at the $r=0$ boundary
of spacetime.  If $\epsilon=0$, this metric describes a portion of
five-dimensional
Anti de Sitter space, with negative cosmological constant.  We could
get more of Anti de Sitter space by continuing to $r=-\infty$.    According
to the AdS/CFT correspondence, quantum gravity
in this five-dimensional spacetime (with $-\infty<r<\infty$)
is equivalent to an ordinary four-dimensional
conformal field theory, without gravity.  The four-dimensional world is
the virtual
``boundary'' at $r=-\infty$.  Randall and Sundrum truncated the Anti de Sitter
space to $r\geq 0$, and observed that in this case one does get
dynamical four-dimensional gravity.\foot{To be more precise, one
gets four-dimensional gravity
coupled to a cutoff version
of the conformal field theory; the low energy physics is conventional
four-dimensional physics with General Relativity coupled to some
massless matter fields that are governed by a nontrivial infrared
fixed point.  The cusp-like behavior of the metric for $r=+\infty$
corresponds in the 
AdS/CFT correspondence to the
nontriviality of the IR fixed point.}  In particular, if $\epsilon=0$,
a four-dimensional graviton is ``localized'' near $r=0$. 

GRS modified this scenario by taking $\epsilon$ nonzero but tiny.
As               long as $e^{-2r/r_0}>>\epsilon$, the correction is
unimportant.  This bound on $r$ means that if the four-dimensional
lengths are not too large, the physics will look effectively four-dimensional,
just as if $\epsilon=0$.  However, if we take $r\to \infty$,  the $\epsilon$
term dominates and the metric becomes five-dimensional Minkowski space.
Hence, at very long distances, the world is effectively five-dimensional.
(I have parametrized the correction in a way that makes it appear
that extreme fine-tuning is required to get $\epsilon$ sufficiently
small.  GRS present the discussion in a way that makes this look
more plausible, though their analysis has a ``negative energy'' problem
that I will mention later.)

\nref\alv{E. Alvarez and C. Gomez, ``Geometric Holography, The Renormalization
Group, and the $c$ Theorem,'' Nucl. Phys. {\bf B541} (1999) 441,
hep-th/9807226.}%
\nref\gir{L. Girardello, M. Petrini, M. Porrati, and
A. Zaffaroni, ``Novel Local CFT and Exact Results On
Perturbations Of $N=4$ Super Yang-Mills From ADS
Dynamics,'' JHEP {\bf 12} (1998) 22, hep-th/9810126.}%
\nref\warner{D. Z. Freedman, S. S. Gubser, K. Pilch,
and N. P. Warner, ``Renormalization Group Flows From
Holography - Supersymmetry And A $c$-Theorem,''
hep-th/9904017.}%
One apparent problem with this scenario (apart from possible issues raised
in \gaba) is that it cannot be realized with matter fields that obey
even the weakest form of a positive energy condition that is usually
considered physically acceptable, which is that the stress tensor
$T_{ij}$ obeys $n^in^jT_{ij}\geq 0$ for any lightlike vector $n$.
Indeed, given this energy condition, one can prove a
``holographic $c$-theorem'' in the AdS/CFT correspondence \refs{\alv - \warner}.
The $c$-theorem says that as one flows to large $r$, the effective
five-dimensional cosmological constant can only become more negative.
Any metric that looks like the ultraviolet end of Anti de Sitter space
near $r=0$, and like five-dimensional Minkowski space near $r=+\infty$,
violates this bound.  In fact, a general five-dimensional
metric with four-dimensional Poincar\'e symmetry can be put in the form
\eqn\sono{ds^2=dr^2+e^{2A(r)}\sum_{i=1}^4 dx_idx^i,}
with some function $A(r)$.  As shown in section 4.2 of \warner,
the weak energy condition implies that $A''\leq 0$.  This inequality
makes it impossible to have a metric that behaves qualitatively like \ulmigu\
in the whole range of $r\geq 0$.  This point has been made independently
in \csaki.

GRS in fact incorporated a brane of negative tension in their
construction of the model. This is one way to violate the weak energy condition.
It seems likely that the physics with violation of the weak energy condition
is unstable.  I do not know if, starting
in $D>5$, it is possible to get a metric
of the qualitative form needed for the GRS scenario while also
respecting the     weak energy condition.

\nref\nima{N. Arkani-Hamed, S. Dimopoulos, N. Kaloper, and
R. Sundrum, ``A Small Cosmological Constant From A Large Extra
Dimension,'' hep-th/000197.}%
\nref\silver{S. Kachru, M. Schulz, and E. Silverstein, ``Self-Tuning
Flat Domain Walls In $5-D$ Gravity And String Theory,''
hep/th/0001206, ``Bounds On Curved Domain Walls in 5-D Gravity,''
hep-th/0002121.}%
I will conclude by briefly mentioning some other interesting
recent developments that were omitted in the actual talk at DM2000
for lack of time.
On the one hand, there are recent proposals 
\refs{\nima,\silver}
attempting to
use a five-dimensional solution with a singularity
(as well as a brane on which standard model fields are localized)
to get a small cosmological constant.  The behavior of these
models will depend very much on what assumption is made about 
the physics of the singularity.  Note that one should not 
expect that a generic singularity will turn out to have any
sensible physical interpretation at all.  Elliptic equations
(such as the ones that describe time-independent
solutions of General Relativity) have an abundance of singular
solutions, and only special ones can be given a physical interpretation
at the quantum level.

I will make a digression on this point.  A
familiar
 example of a singular solution that
should {\it not}
be given a physical interpretation is the singular
Dirac monopole solution of QED.  
We do not want to predict magnetic monopoles in QED with a mass
that (in units of the electron mass) would only depend on the
fine structure constant!
By contrast to this example,
all of the singular charge-bearing ``brane'' solutions of 
ten and eleven-dimensional supergravity have turned out to be
approximations to objects that actually exist in the pertinent
quantum theories.  
Apart from the fact that fortune sometimes favors
    the brave, why did this occur?
I think it may be that in the presence of
gravity, since
 a charge-bearing singularity can be hidden behind
a black hole horizon, objects carrying all of the gauge     charges
must exist.  
This contrasts with QED which does not include gravity and does
not have magnetic monopoles.
In any event, the generic singularity of a
time-independent solution of General Relativity is much ``worse''
-- much less likely to have a quantum interpretation -- than
the minimal charge-bearing singularities.

So the singularities encountered in \refs{\nima,\silver}
may well not have any sensible physical interpretation at all,
especially in cases in which (as described in the second paper
in \silver) the singular solution has a modulus that controls
the four-dimensional cosmological constant.  If the solution with
the singularity were physically sensible, one would expect the
modulus to correspond to a physical field, which would then relax
to minimize the vacuum energy -- not necessarily reaching the value zero.

If a singularity
does have a physical interpretation in terms of four-dimensional
effective field theory, perhaps with some fields supported near
the singularity, this still seems likely at best to lead to
a restatement of the problem of the cosmological constant,
in a new context.  The best hope might be that the framework
of low energy effective four-dimensional field theory would somehow
break down because of the behavior near the singularity, perhaps
leading to something roughly along the lines of GRS.

Finally,
another interesting development is the   study of
nonplanar loop diagrams in noncommutative geometry
\ref\minrs{S. Minwalla, M. Von Raamsdonk and N. Seiberg,
``Noncommutative Perturbative Dynamics,'' hep-th/9912072.}.
A surprising connection between infrared and ultraviolet dynamics
was found, wherein what would usually be regarded as the
ultraviolet region of a Feynman diagram produces infrared singularities.
This evades the usual renormalization group thinking by means of
which one usually deduces an
 effective low energy field theory description -- which leads
to trouble with the cosmological constant.
(However, the specific phenomenon found in \minrs\ can be described
in a low energy effective field theory, albeit one that contains
fields that one might not expect.)
As has been noted by the authors of \minrs, such an IR/UV
connection might, if 
other elements fall into place, be an
  ingredient in an eventual solution of the
cosmological constant problem.  

\bigskip
This work was supported in part by NSF Grant PHY-9513835 and the
Caltech Discovery Fund.
\listrefs
\end